\documentclass[manuscript]{aastex63}
\usepackage{amsmath}

\shorttitle{Magnetic Twists of Solar Filaments}

\begin{document}

\title{Magnetic Twists of Solar Filaments}

\correspondingauthor{Y. Guo and P. F. Chen}
\email{guoyang@nju.edu.cn, chenpf@nju.edu.cn}

\author[0000-0002-4205-5566]{J. H. Guo}
\affiliation{School of Astronomy and Space Science, Nanjing University, Nanjing 210023, China}
\affiliation{Key Laboratory of Modern Astronomy and Astrophysics (Nanjing University), Ministry of Education, Nanjing 210023, China}

\author[0000-0002-9908-291X]{Y. W. Ni}
\affiliation{School of Astronomy and Space Science, Nanjing University, Nanjing 210023, China}
\affiliation{Key Laboratory of Modern Astronomy and Astrophysics (Nanjing University), Ministry of Education, Nanjing 210023, China}

\author[0000-0002-1190-0173]{Y. Qiu}
\affiliation{School of Astronomy and Space Science, Nanjing University, Nanjing 210023, China}
\affiliation{Key Laboratory of Modern Astronomy and Astrophysics (Nanjing University), Ministry of Education, Nanjing 210023, China}

\author[0000-0001-5483-6047]{Z. Zhong}
\affiliation{School of Astronomy and Space Science, Nanjing University, Nanjing 210023, China}
\affiliation{Key Laboratory of Modern Astronomy and Astrophysics (Nanjing University), Ministry of Education, Nanjing 210023, China}

\author[0000-0002-9293-8439]{Y. Guo}
\affiliation{School of Astronomy and Space Science, Nanjing University, Nanjing 210023, China}
\affiliation{Key Laboratory of Modern Astronomy and Astrophysics (Nanjing University), Ministry of Education, Nanjing 210023, China}

\author[0000-0002-7289-642X]{P. F. Chen}
\affiliation{School of Astronomy and Space Science, Nanjing University, Nanjing 210023, China}
\affiliation{Key Laboratory of Modern Astronomy and Astrophysics (Nanjing University), Ministry of Education, Nanjing 210023, China}

\begin{abstract}
Solar filaments are cold and dense materials situated in magnetic dips, which show distinct radiation characteristics compared to the surrounding coronal plasma. They are associated with coronal sheared and twisted magnetic field lines. However, the exact magnetic configuration supporting a filament material is not easy to be ascertained because of the absence of routine observations of the magnetic field inside filaments. Since many filaments lie above weak-field regions, it is nearly impossible to extrapolate their coronal magnetic structures by applying the traditional methods to noisy photospheric magnetograms, in particular the horizontal components. In this paper, we construct magnetic structures for some filaments with the regularized Biot--Savart laws and calculate their magnetic twists. Moreover, we make a parameter survey for the flux ropes of the Titov-D\'emoulin-modified model to explore the factors affecting the twist of a force-free magnetic flux rope. It is found that the twist of a force-free flux rope, $|\overline{T_{\rm w}}|$, is proportional to its axial length to minor radius ratio $L/a$, and is basically independent of the overlying background magnetic field strength. Thus, we infer that long quiescent filaments are likely to be supported by more twisted flux ropes than short active-region filaments, which is consistent with observations.
\end{abstract}

\keywords{Sun: solar corona --- Sun: filaments, prominences --- Sun: solar magnetic fields}

\section{Introduction} \label{sec:intro}

Solar filaments are one of the most fascinating structures embedded in the corona. Their temperature ($\sim$7000 K) is much lower and their density ($10^{10}$--$10^{11}\ \rm cm^{-3}$) is much higher than the surroundings, causing distinct radiative characteristics. They usually appear as dark and extended filamentary structures when viewed against the solar disk, while are seen as bright cloud-like prominences when seen beyond the solar limb. It has been revealed that the cold and dense plasmas originate from the chromosphere \citep{Spicer98, Wang1999, Berger11, Xia16}. Solar filaments are intimately related to solar flares and coronal mass ejections \citep[CMEs,][]{Chen2011}, which are powered by nonpotential coronal magnetic field. Therefore, solar filaments are the effective tracers of sheared and/or twisted magnetic structures. Generally, the magnetic structure of filaments is described by two typical models: the inverse-polarity magnetic flux rope (MFR) model \citep{Kuperus74}, and the normal-polarity magnetic sheared arcade model \citep{Kippenhahn57}. \citet{chen14} proposed an indirect method to determine the magnetic configuration type of a filament based on the conjugate draining sites upon filament eruption. Applying this method to 571 solar filaments, \citet{Ouyang15, Ouyang17} found that 89\% of solar filaments are supported by MFRs, and 11\% are supported by magnetic sheared arcades.

Based on different magnetic environments, it is quite general to divide filaments into quiescent and active-region types \citep{Mackay2010}. Many previous works have revealed that these two types of filaments have distinct properties, such as the morphology, dynamics, magnetic configuration, and so on \citep{McCauley15, Ouyang17, Xing18, Zou19}. Regarding the morphology, quiescent filaments are typically 30 Mm high and 200 Mm long, whereas active-region filaments are usually less than 10 Mm high and typically 50 Mm long \citep{Tanberg-Hanssen1974, Tanberg1995, Filippov2000, Engvold2015}. Although overall 89\% of solar filaments are supported by MFRs, the preference of magnetic configuration of two types of filaments is different: Regarding quiescent filaments, 96\% of them are supported by MFRs, only 4\% are supported by sheared arcades. On the contrary, for active-region filaments, 40\% of them are supported by sheared arcades \citep{Ouyang17}. This result seems counterintuitive since, compared to quiet regions, active regions are more non-potential and more complex, which favors the existence of an MFR \citep{Ballehooijen89, Mackay08, Xia14b}. In this paper, we attempt to provide an explanation for this puzzle from the perspective of magnetic topology.

Coronal magnetic field is essential to fully understand the filament structures. However, the magnetic field in the corona is difficult to be measured directly so far. Therefore, several approaches have been proposed to diagnose the coronal magnetic field. One method is coronal seismology, which utilizes coronal loop oscillations or filament oscillations \citep{Tripathi09, zhang13}. Recently, \citet{Yang2020} mapped the global magnetic field in the corona based on the magnetoseismology. Another extensively used method is nonlinear force-free field (NLFFF) modeling \citep{Yan01, Regnier04, Canou09, Guo10, Jing10, Inoue10, Jiang14, Yang15, Yang16, Zhong19, Qiu20}. However, when the photospheric magnetic field is weak or the flux rope is high-lying, the twisted magnetic structure is hard to be obtained by the traditional extrapolation methods. This is because the traditional extrapolation methods strongly rely on the photospheric magnetogram, which is the only constraint. If the horizontal magnetic field is not significantly larger than the noise, the extrapolation frequently leads to a simpler magnetic configuration in the corona, i.e., low-lying sheared arcades. In addition, information is hard to be transformed correctly from the bottom to high-lying regions due to numerical errors. Unfortunately, many filaments are located in decayed active regions or quiet regions, where the horizontal magnetic field is very weak. To construct complicated magnetic structures with a possible MFR, some researchers suggested that coronal observations can be utilized to constrain the magnetic structure. For example, \citet{Ballegooijen04} proposed the flux rope insertion method to construct a twisted magnetic structure directly, which has been extensively applied in many events \citep{Bobra08, Savcheva09, Su15, Huang19}. It is noted that the inserted flux rope deviates from an equilibrium state, which can be achieved by magneto-frictional relaxation. Based on this method, some magnetic structures of quiescent or polar crown filaments have been constructed \citep{Su12, Su15, Luna17, Mackay20}. Recently, \citet{Titov18} proposed to use the regularized Biot--Savart laws (RBSLs) to model an approximately force-free flux rope. One of the advantages of this technique is that the flux rope can have any shape in the three-dimensional space. With this technique, \citet{Guo19} successfully constructed an MFR for a large-scale filament with weak horizontal magnetic field on the solar surface.

In this paper, we construct the twisted magnetic structures for solar filaments and explore what factors affect their twist. Our work has three aims: (1) to construct the twisted magnetic structures for solar filaments with the RBSL method, (2) to understand the factors that affect the twist of an MFR, and (3) to explain the magnetic configuration preference between active-region filaments and quiescent filaments. The paper is organized as follows. In Section \ref{sec:obs}, we construct magnetic structures for five solar filaments with the RBSL method constrained by observations. In Section \ref{sec:the}, we explore the factors affecting the twist of an MFR using the Titov-D\'emoulin-modified (TDm) model. The summary and discussions are given in Section \ref{sec:dis}.

\section{Magnetic Structures in Observations} \label{sec:obs}

\subsection{Coronal magnetic field reconstruction} \label{sec:obs1}

We reconstruct the magnetic structures of five filaments observed by the \emph{Solar Dynamics Observatory} \citep[\emph{SDO};][]{Pesnell12} and by the \emph{Solar Terrestrial Relations Observatory} (\emph{STEREO}) missions. The Atmospheric Imaging Assembly \citep[AIA;][]{Lemen12} on board \emph{SDO} provides full-disk coronal images at extreme ultraviolet (EUV) wavebands. Its pixel size is $0\farcs6$ and temporal cadence is 12 s. Meanwhile, the Extreme Ultraviolet Imager \citep[EUVI;][]{Wuelser04} of the Sun Earth Connection Coronal and Heliospheric Investigation \citep[SECCHI;][]{Howard08} suite on board \emph{STEREO} supplies other views of the Sun at similar wavelengths, such as 171 and 304 \AA. Thus, similar to previous works \citep{Zhou17, Zhou19, Guo19, Xu20}, we can construct the three-dimensional filament geometry by the triangulation technique, and take it as the main axis of the MFR. The vector magnetic field is obtained by the Helioseismic and Magnetic Imager \citep[HMI,][]{Scherrer12, Schou12, Hoeksema14} on board \emph{SDO}. Similar to \citet{Titov18} and \citet{Guo19}, we construct the MFR with the RBSL method, and embed it in a potential magnetic field, which is extrapolated with the Green's function method \citep{Chiu77}. Then, the newly combined magnetic field is relaxed to a force-free state by the magneto-frictional code \citep{Guo2016a, Guo2016b}. The aforementioned reconstructions are performed with the Message Passing Interface Adaptive Mesh Refinement Versatile Advection Code \citep[MPI-AMRVAC;][]{Keppens12, Xia18, Keppens20}. We select five eruptive filaments, which are all observed by \emph{SDO} and \emph{STEREO} simultaneously (see Table \ref{tab1}). Taking the first event as an example, the reconstruction process is described in detail in the following two paragraphs.

\begin{figure}[ht!]
\centering
\includegraphics[scale=0.40]{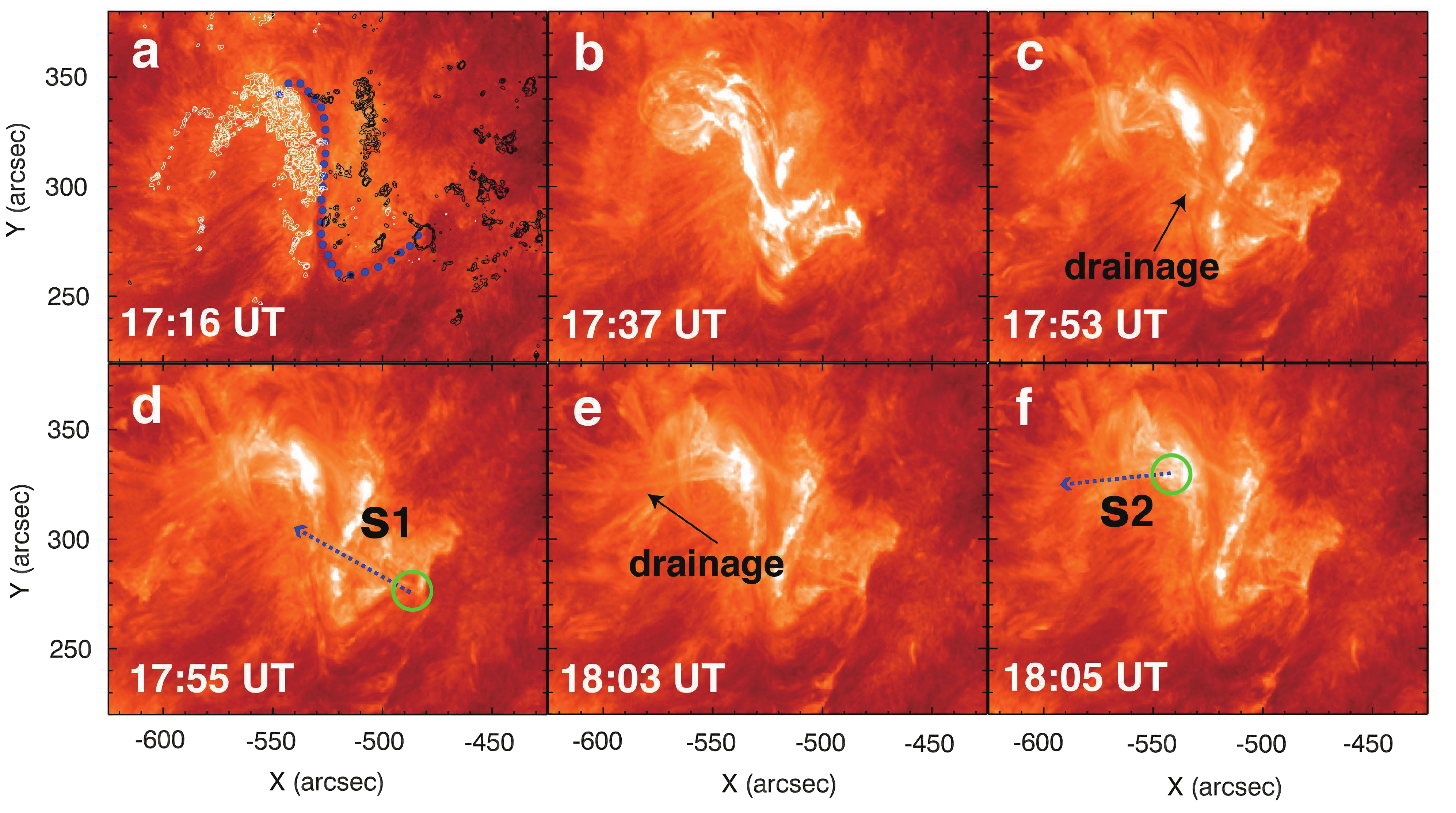}
\caption{AIA 304 \AA \ images showing the filament eruption process on 2012 May 5 at different moments. (a) The filament is outlined by blue solid circles. The white/black contour lines correspond to the positive/negative magnetic polarities. (b) The filament rises and then rotates counterclockwise. (c)--(f) Filament material slides down along the field lines. The arrows in panels (c) and (e) mark the draining sites. The green circles in panels (d) and (f) denote the conjugate draining sites of the filament. The blue dotted lines along the material draining track mark the slice positions used for Figure \ref{fig2}, where the arrows denote the directions of the slices.}
\label{fig1}
\end{figure}

\begin{figure}[ht!]
\centering
\includegraphics[scale=0.88]{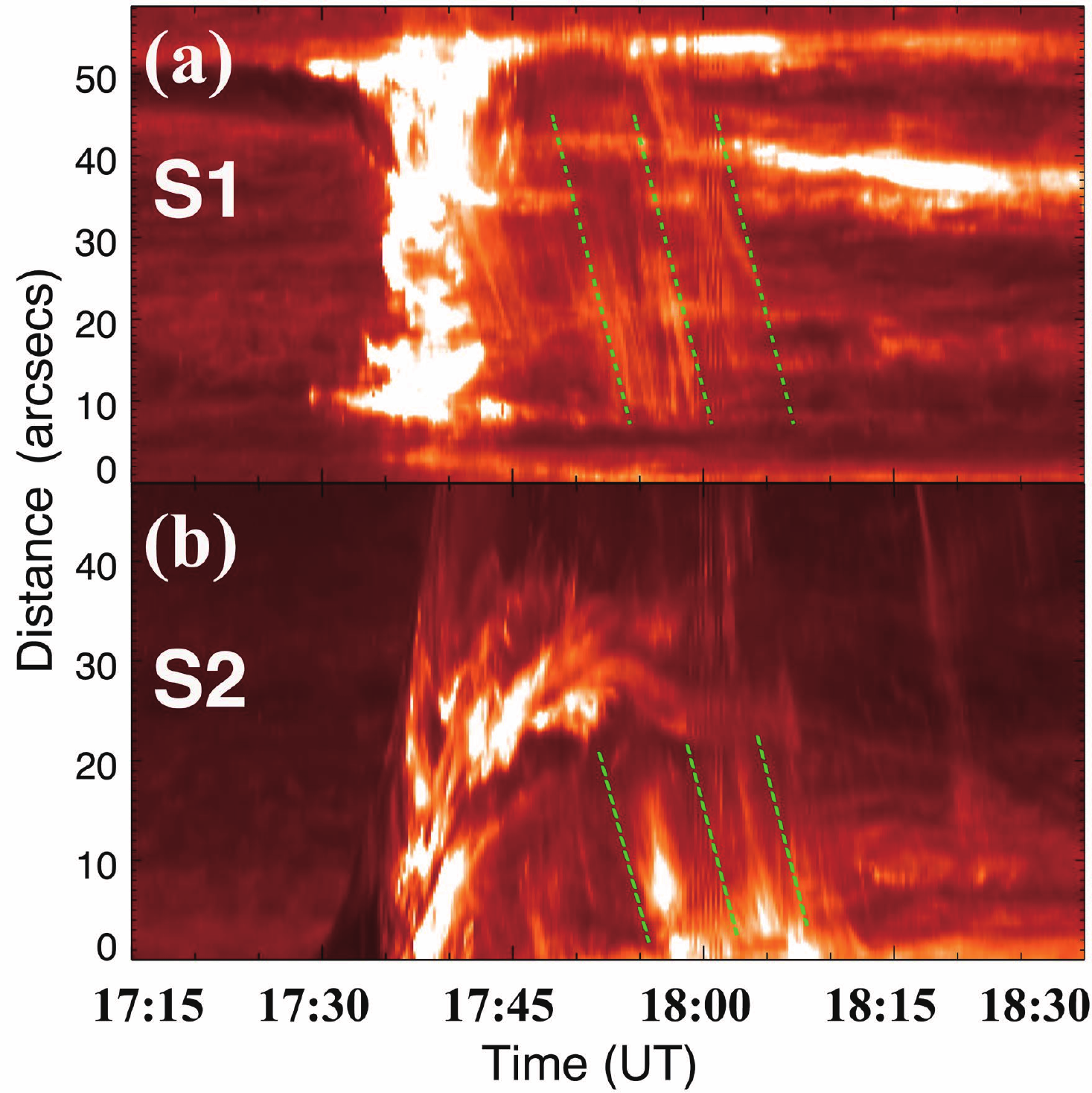}
\caption{Time-distance diagrams of the filament material along slices S1 (a) and S2 (b) as shown in Figure \ref{fig1}, where the green dashed lines indicate the filament material draining motions.}
\label{fig2}
\end{figure}

First, we need to construct the 3D path of the MFR axis. In the previous works \citep{Zhou17, Zhou19, Guo19, Xu20}, the triangulation method was adopted to construct the 3D path of the filament axis, which is taken as the MFR axis. However, this method does not work well for the low altitude parts of a filament. Moreover, the endpoints of a filament might not correspond to the footpoints of the supporting flux rope \citep{hao16, chen20}. Thus, we propose a new method to approximate the 3D path of the MFR axis. According to \citet{chen14}, the drainage sites upon filament eruption shed light on the MFR footprints. For example, the failed eruption of an active-region filament is observed by \emph{SDO}/AIA 304 \AA \ on 2012 May 5, as illustrated in Figure \ref{fig1}. At about 17:20 UT, this filament starts to rise and then rotates counterclockwise. Subsequently, some filament material slides down along the field lines as manifested by the new threads, which are pointed to by the black arrows in Figures \ref{fig1}c and \ref{fig1}e. The material draining forms two conjugate weakly bright spots in the lower atmosphere, as indicated by the green circles in Figures \ref{fig1}d and \ref{fig1}f. The drainage sites are left-skewed relative to the magnetic polarity inversion line, implying this filament is dextral and corresponds to left-handed (i.e., with a negative sign) magnetic helicity \citep{Wang09,chen14}. To see the material draining dynamics clearly, we select two slices along the material draining track, as indicated by the blue dotted lines in Figures \ref{fig1}d and \ref{fig1}f. The time-distance diagrams of the AIA 304 \AA \ intensity are shown in Figure \ref{fig2}, and one can see that the filament material moves downward. Once the MFR footprints are determined, the projection path of the MFR axis is outlined by blue solid circles in Figure \ref{fig1}a. We adopt the triangulation method to measure the height of the filament axis, where the apex of the filament is recorded, but the heights of the rest points are fitted by a semicircle arc linking the filament apex and each MFR footpoint. Finally, since the RBSL method requires a closed path, we close it by adding a mirror sub-photosphere arc. So far, we have obtained the 3D information of the MFR axis.

\begin{figure}[ht!]
\centering
\includegraphics[scale=0.6]{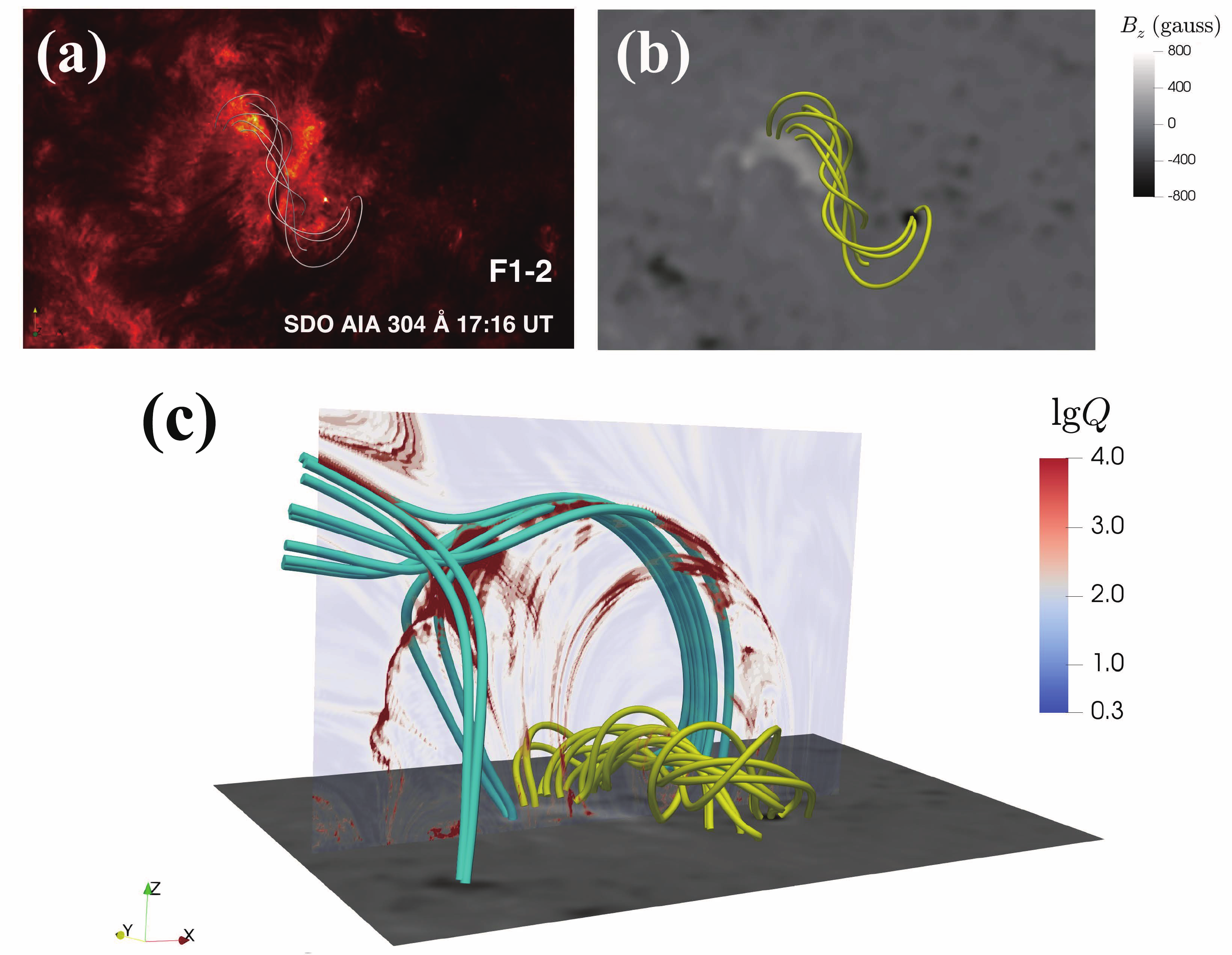}
\caption{(a) AIA 304 \AA \ image of the filament in event F1, where the white lines denote magnetic field lines of the MFR. (b) Longitudinal magnetogram (grey-scale) superposed by the extrapolated coronal magnetic field (yellow lines). (c) Side view of the extrapolated coronal magnetic field lines (cyan and yellow lines) and the vertical slice shows distribution of the corresponding squashing factor $Q$ (color scale).}
\label{fig3}
\end{figure}

\begin{figure}[ht!]
\centering
\includegraphics[scale=0.36]{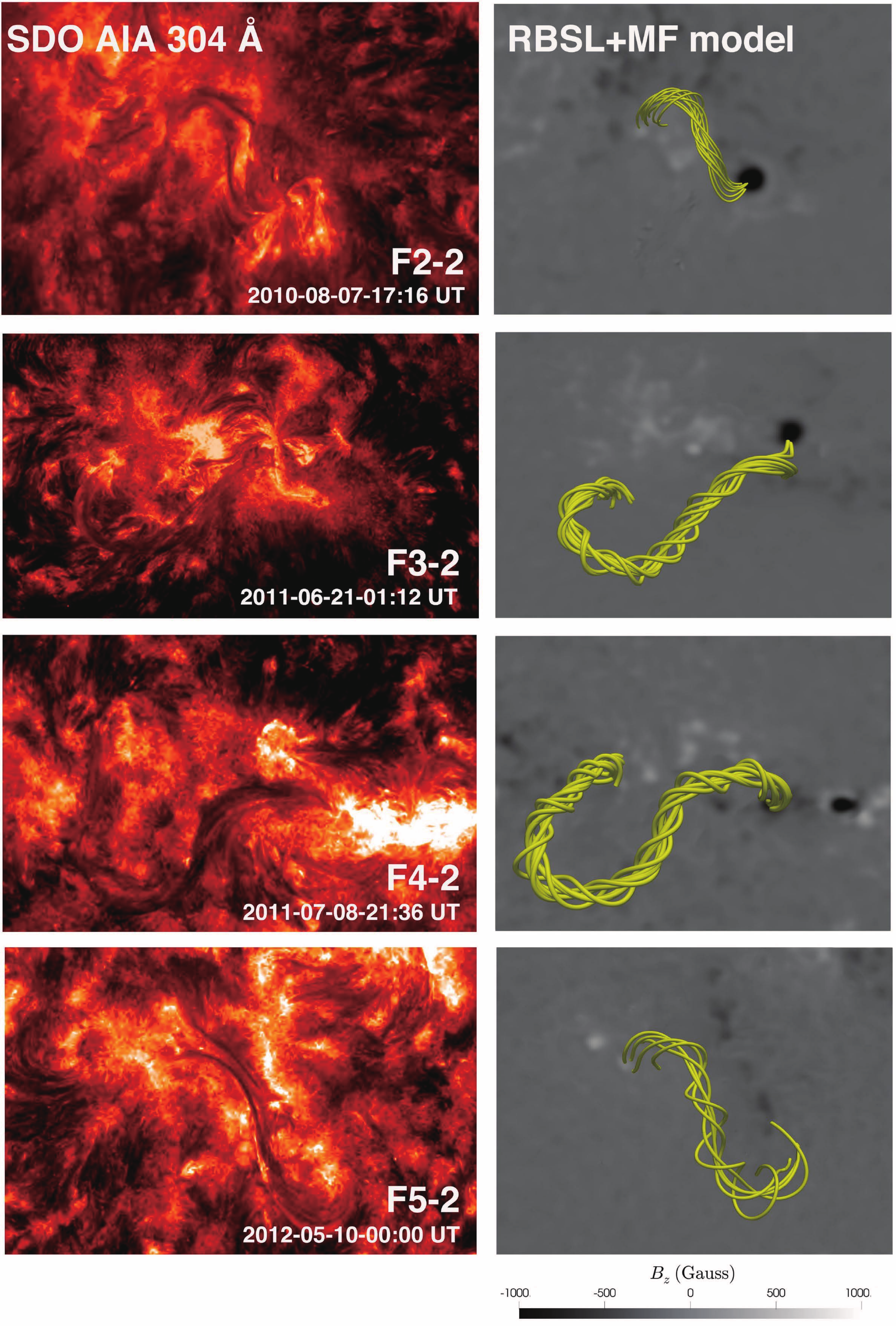}
\caption{AIA 304 \AA \ images (left column) and the extrapolated coronal magnetic field lines superposed on longitudinal magnetograms (right column) of other events.}
\label{fig4}
\end{figure}

Second, we construct the magnetic structure of the filament. It contains two steps, namely vector magnetic field pretreatment and coronal magnetic field extrapolation. With respect to the vector magnetic field pretreatment, we need to correct the projection effect \citep{Guo17b}, and to remove the net Lorentz force and torque \citep{Wiegelmann06}. Regarding the coronal magnetic field extrapolation, we utilize the RBSL method, where we need to set the physical parameters at first, i.e., the minor radius of an MFR ($a$), the axial magnetic flux ($F$), and the electric current ($I$) in addition to the 3D path of the MFR axis we obtained in the previous paragraph. However, the minor radius of the flux rope, $a$, is hard to measure directly, so we set it as a free parameter, and two values are chosen for each event in Table \ref{tab1}. Accordingly, the magnetic flux $F_0=(|F_+| +|F_-|)/2=4.13 \times 10^{20}\ \rm Mx$ within the minor radius $a$ can be estimated with the vertical component of the field. In this event, we take $F=5F_0$ after a few tests since the MFR is likely to disappear during the ensuing relaxation if the magnetic flux $F$ is too small. Moreover, the axial flux $F$ is only an initial parameter for the RBSL method. It has to be added to the background potential field in order to maintain the vertical magnetic field in observations. Then, the current $I$ is determined by Equation (12) in \citet{Titov18}. Next, we embed the constructed MFR into the background potential field and relax the aforementioned model to a force-free state by the magneto-frictional method \citep{Guo2016a, Guo2016b}. After relaxation, the force-free metric is $\sigma_{J}=0.20$, and the divergence-free metric is $\langle |f_{i}|\rangle =3.08 \times 10^{-4}$ \citep[see][for details of the two metrics]{Guo2016b}. Compared to previous studies, the values of these two metrics are basically acceptable \citep{Guo2016a, Guo19, Zhong19}. The reconstructed 3D magnetic field after relaxation is shown in Figure \ref{fig3}. One can see clearly that the reconstructed MFR shows a similar shape compared to the filament in the 304 \AA \ observation. The reconstruction results of other events are shown in Figure \ref{fig4}.

\subsection{Magnetic Topology \label{sec:obs2}}

\begin{figure}[ht!]
\centering
\includegraphics[scale=0.9]{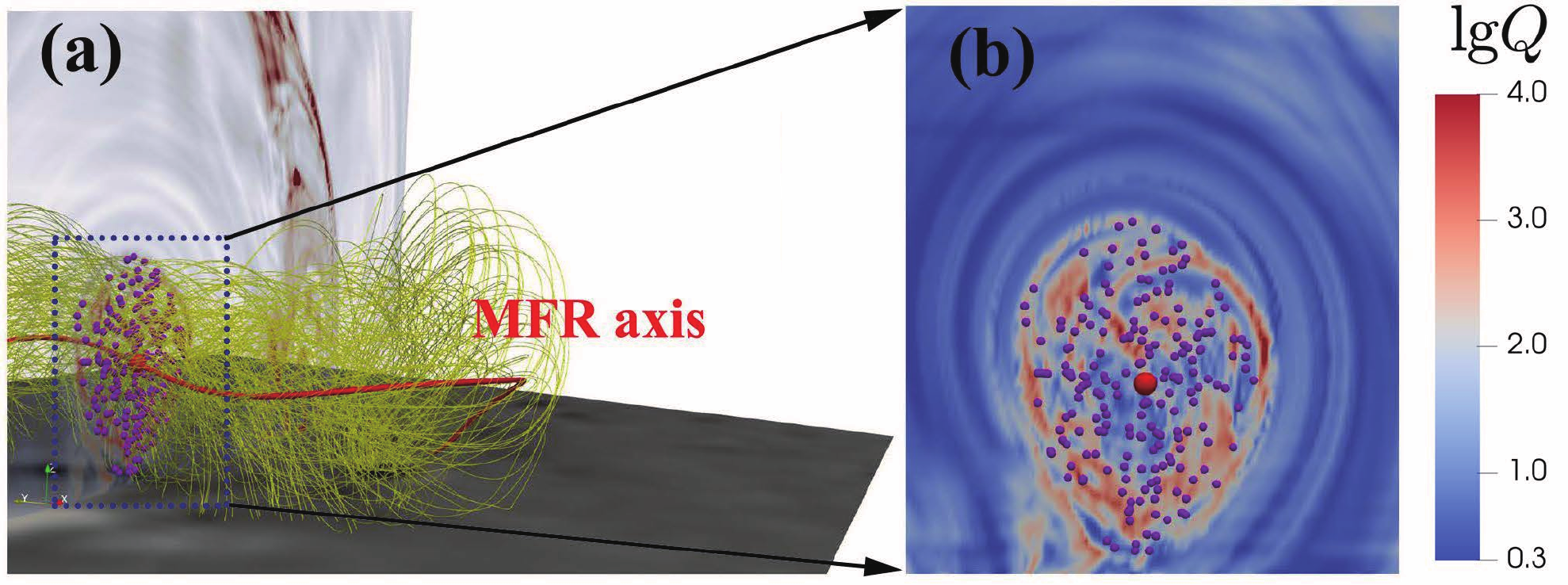}
\caption{(a) Distribution of the squashing factor $Q$ (color scale) and sample magnetic field lines (yellow lines) superposed on the longitudinal magnetogram (grey scale). The red line denotes the MFR axis. (b) Zoom-in view of the $Q$-map in the $xz$-plane at $y=0$. The red dot indicates the intersection of the MFR axis with the $xz$-plane and the purple dots indicate the intersections of the sample field lines of the MFR with the $xz$-plane.}
\label{fig5}
\end{figure}

\begin{figure}[ht!]
\centering
\includegraphics[scale=0.29]{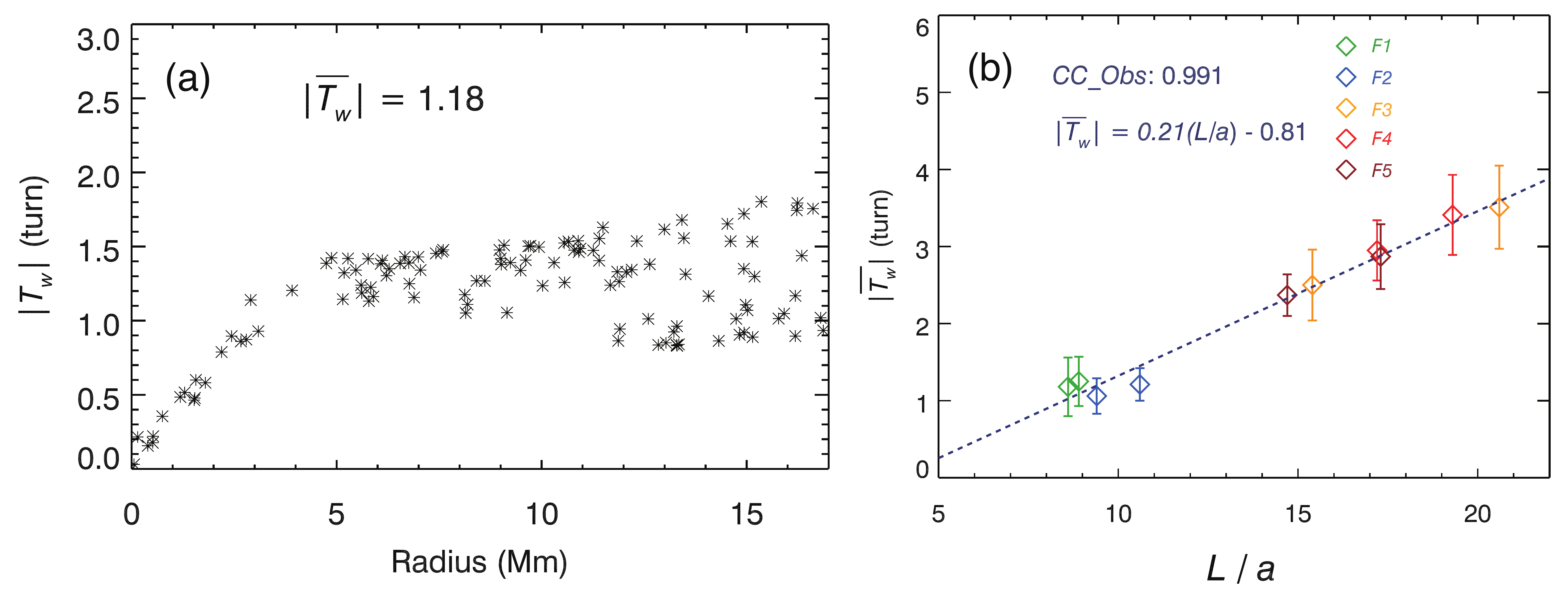}
\caption{(a) Variation of the twists of selected magnetic field lines with the distance from the MFR axis in F1-2. (b) Mean twist of MFRs as a function of their aspect ratio $L/a$ of the five filaments as listed in Table \ref{tab1}. The data points of an individual filament are denoted by the same color. The dashed line is the linear fitting to the data points.}
\label{fig6}
\end{figure}

To further understand the physical properties of MFRs, we compute the magnetic topological parameters including the quasi-separatrix layers (QSLs) and the twists. QSLs can be determined by calculating the squashing factor $Q$ \citep{Priest95, Titov02}. We calculate $Q$ using the method proposed by \citet{Scott17}, but with code written by Kai E. Yang \footnote{https://github.com/Kai-E-Yang/QSL}. QSLs depict the border of an MFR and the favorite places for magnetic reconnection \citep{Guo13, Guo17, Liu16}. So, we measure the minor radius of an MFR by QSLs. The $Q$ distribution in the $xz$-plane at $y=0$ is shown in Figures \ref{fig3}c and \ref{fig5}a, which clearly shows that the flux rope is surrounded by a quasi-circular QSL, and the minor radius of the MFR is estimated to be about 17 Mm. Then, we calculate the twist number in the following three steps following the formula given by \citet{Berger06}, which was implemented by \citet{Guo17}. First, we cut a slice almost perpendicular to the MFR. Second, we select the magnetic field line that is most perpendicular to the slice and take it as the main axis, as denoted by the red line in Figure \ref{fig5}a. Then, two hundred sample magnetic field lines within the MFR are selected to calculate the mean twist, as denoted by the yellow lines in Figure \ref{fig5}a. Figure \ref{fig5}b illustrates the intersections of the MFR axis (red dot) and the sample field lines (purple dots) within the $xz$-plane. We find that the mean twist for this case is about -1.18 turns, and the standard deviation is about 0.38. Figure \ref{fig6}a displays the twist numbers of the selected magnetic field lines at different distances from the MFR axis. Clearly, the MFR core has a lower twist than the outer shell.

In the same way, we also reconstruct the magnetic structures for the other four filaments. The geometric parameters of the MFRs and their mean twist numbers are listed in Table \ref{tab1}. Apparently the mean twist of an MFR is proportional to the axis length, $L$, and inversely proportional to its minor radius, $a$, defined by the quasi-circular QSL. In Figure \ref{fig6}b, we plot the mean twist numbers of the five filaments as a function of $L/a$, the aspect ratio of the MFR. Note that for each filament there are two data points since two values of $a$ are assumed for each filament. It is seen that the mean twist almost linearly varies with $L/a$, with a correlation coefficient of 0.991. The inferred least-squares best fit is $|\overline{T_{\rm w}}|=0.21(L/a)-0.81$, as shown in Figure \ref{fig6}b. It seems that, for a flux rope constructed by the RBSL method, the twist amount is highly related to its geometric structure.

\section{Theoretical model} \label{sec:the}

\subsection{Construction of flux ropes based on the TDm model} \label{sec:the1}

In Section \ref{sec:obs}, we constructed MFRs for five filaments, and found a quantitative relationship between the twist of an MFR and its geometrical parameter $L/a$. However, it is not easy to perform a large parameter survey through observations. Instead, it would be helpful if we can check the correlation with a theoretical model. It is noticed that the TDm model provides analytical solutions for force-free MFRs including two kinds of electric current profiles \citep{Titov14}. One is concentrated in a thin layer at the boundary of the MFR, and the other has a parabolic profile, decreasing from the MFR axis to the boundary. In this section, we adopt the parabolic profile for the current density, and then construct the TDm model with the RBSL method \citep{Titov18}. The procedure is as follows.

First, we construct the bipolar-type background magnetic field $\boldsymbol{B_{\rm q}}$. We set two sub-photosphere magnetic charges of strength $q$ at a depth of $d_{\rm q}$, which lie on the $y$-axis at $y=\pm L_{\rm q}$. Second, we set physical parameters for the RBSL method. Different from that in Section \ref{sec:obs}, the axis path of an MFR in the TDm model has a semicircular shape, which follows a contour line of the potential magnetic field $B_{\perp}$ on the $y=0$ plane. The axis also needs to be closed by adding a sub-photosphere semicircle. Hence, the flux-rope axis is a circular arc with a major radius $R_c$. According to the force balance conditions, the electric current $I$ is determined by Equation (7) in \citet{Titov14}, and the axial flux of the MFR is calculated by Equation (12) in \citet{Titov18}. Third, we embed the flux rope constructed by the RBSL method with the aforementioned parameters into the potential magnetic field. Finally, we relax the combined magnetic field to a quasi-force-free state by the magneto-frictional code \citep{Guo2016a, Guo2016b}. As the benchmark, we set $q=100 \ \rm T \ Mm^{2}$, $L_{\rm q}=50\rm \ Mm$, $d_{\rm q}=50\rm \ Mm$, $R_{\rm c}=94.2\rm \ Mm$ and $a=20 \rm \ Mm$, which is labeled as case RF, standing for ``reference". After the relaxation, the force-free metric is $\sigma_{J}=0.086$, and the divergence-free metric is $\langle |f_{i}|\rangle=1.47 \times 10^{-5}$, indicating the flux rope model is almost force-free and divergence-free. Then we measure the minor radius of the resulting MFR and calculate its twist number similar to those in Section \ref{sec:obs}. It is found that the mean twist is about $2.27\pm 0.98$ turns.

\subsection{Parameter survey \label{sec:the2}}

\begin{figure}[ht!]
\centering
\includegraphics[scale=0.52]{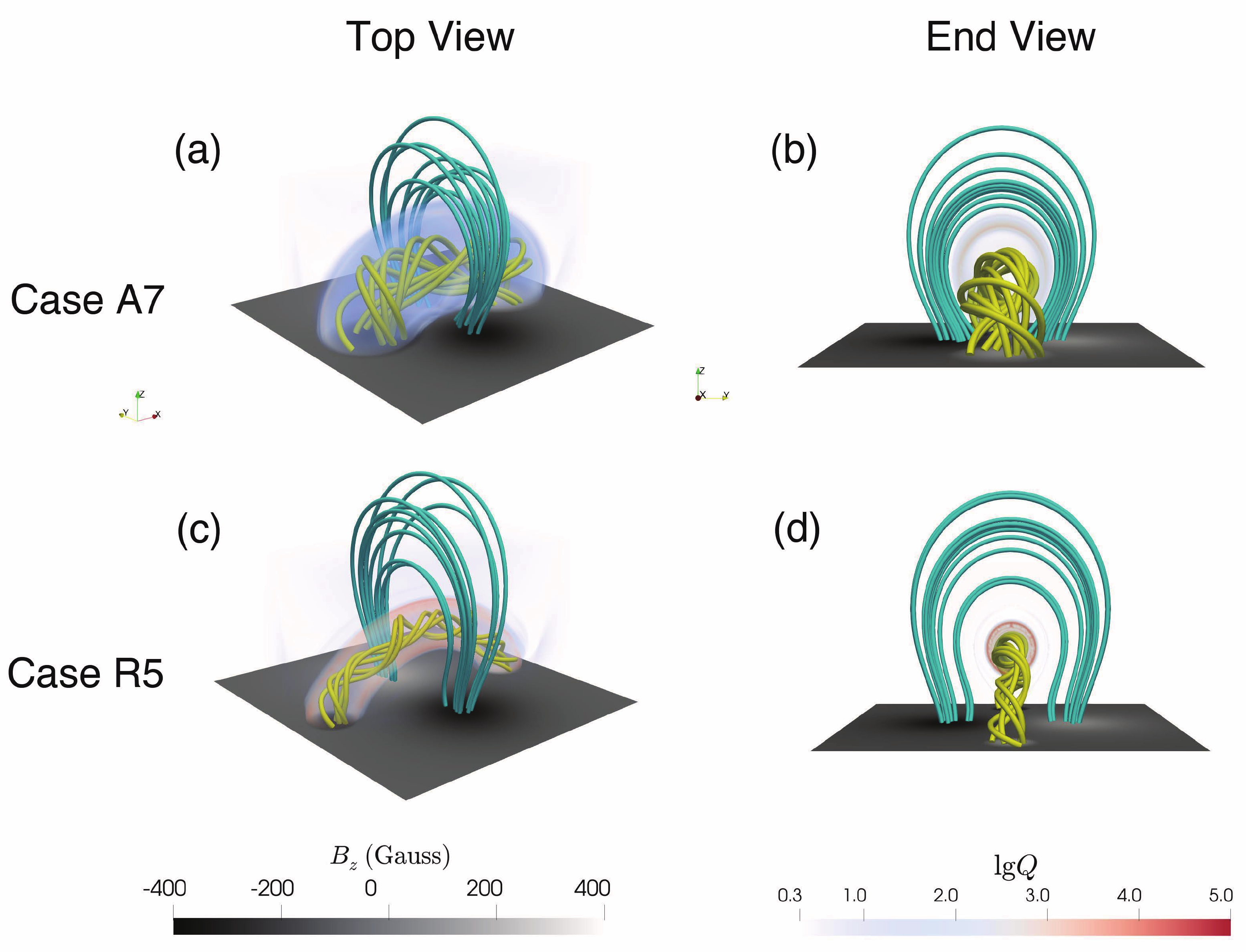}
\caption{Magnetic field lines and the QSLs in cases A7 (top row) and R5 (bottom row). Panels (a) and (c) correspond to the top view, and panels (b) and (d) correspond to the end view along the main axis of the flux rope. The cyan lines denote the overlying potential field, whereas the yellow lines denote the magnetic field of the MFRs. Semi-transparent isosurfaces delineate the QSLs.}
\label{fig7}
\end{figure}

We explore how the twist of an MFR depends on its geometric and magnetic parameters, including the MFR length ($L$), the MFR minor radius ($a$), and the background magnetic field strength ($B_{\rm q}$). To explore the effect of $L$, we construct six magnetic field models with different $R_{\rm c}$ as listed in Table \ref{tab2}, and the other input parameters remain the same as those in case RF. Similarly, to investigate the effect of the MFR minor radius, we choose eight values of $a$ as listed in Table \ref{tab2}, with the other input parameters being the same as those in case RF. Besides, we also study the influence of the background magnetic field strength on the twist of the resulting MFR. Keeping the other input parameters being the same as those in case RF, we set the magnetic charges with different strengths. Table \ref{tab2} lists the input parameters ($R_{\rm c},\ L,\ a,\ L/a,\ q,\ d_{\rm q},\ L_{\rm q}$) and the output quantities ($|\overline{T_{\rm w}}|,\ \sigma_{J},\ \langle |f_{i}|\rangle$) of the MFRs, where all the cases are categorized into three groups, with $R_{\rm c}$, $a$ and $q$ being the control parameters individually. Figure \ref{fig7} depicts the QSLs and selected magnetic field lines in cases A7 and R5. It shows clearly that the MFRs are wrapped by closed QSLs.

\begin{figure}[ht!]
\centering
\includegraphics[scale=0.4]{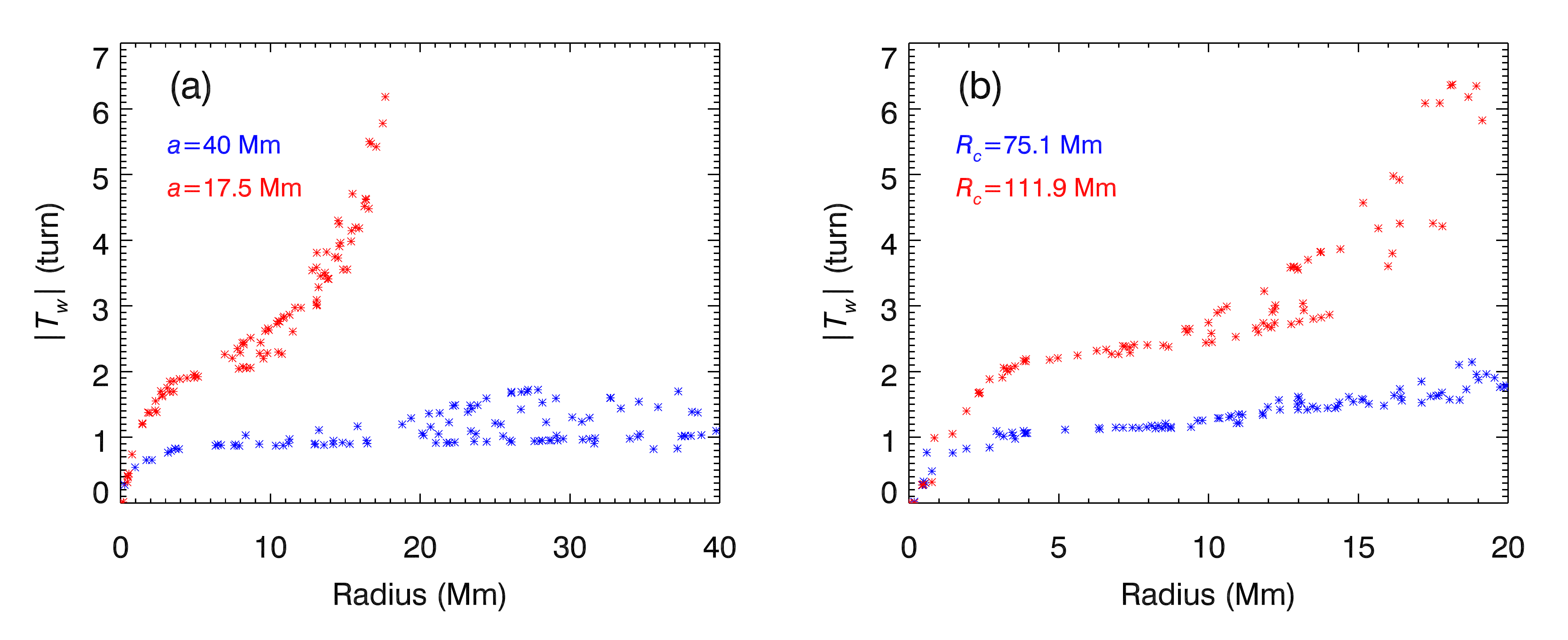}
\caption{Variation of the twist of selected magnetic field lines with the distance from the MFR axis. (a) Results in cases A7 (blue asterisks) and A3 (red asterisks) with different minor radii of the MFR. (b) Results in cases R1 (blue asterisks) and R5 (red asterisks) with different major radii of the MFR.}
\label{fig8}
\end{figure}

\begin{figure}[ht!]
\centering
\includegraphics[scale=0.5]{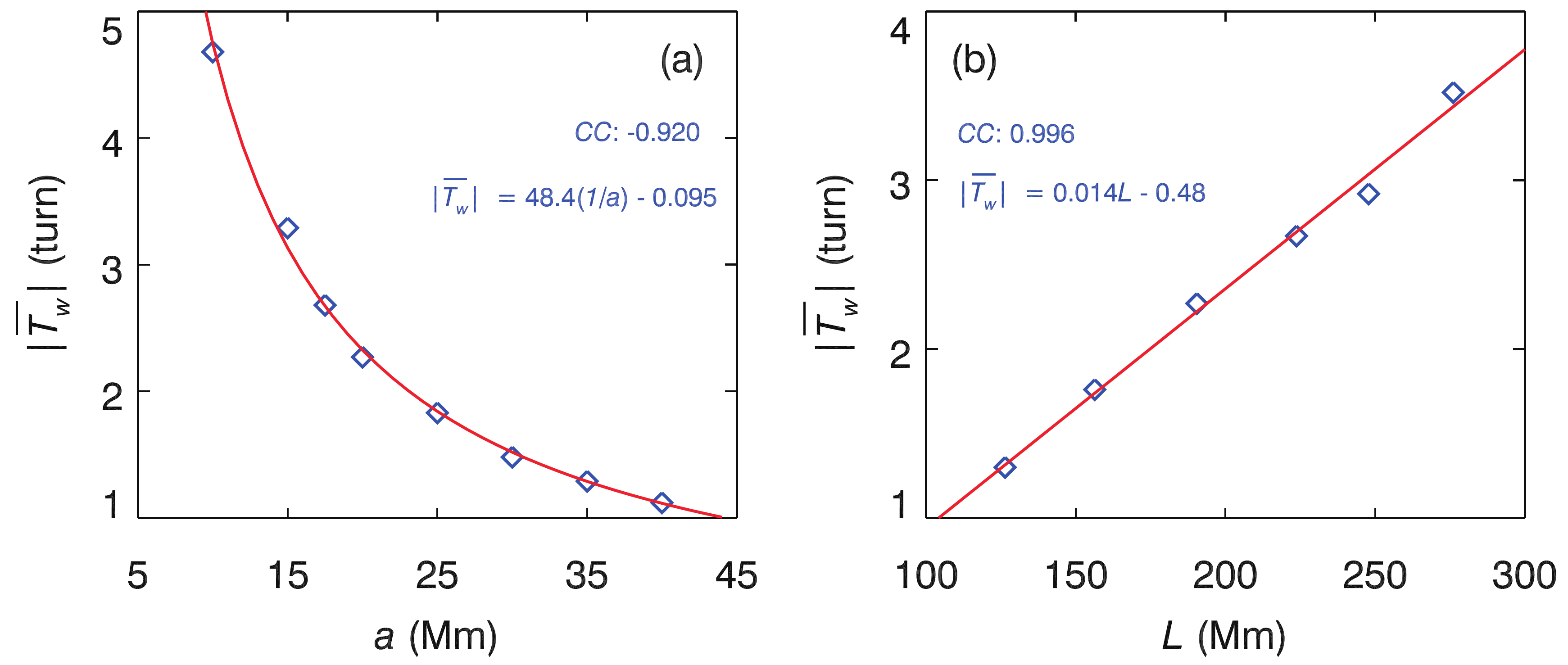}
\caption{Variation of the absolute value of the mean twist of an MFR, ${|\overline{T_{\rm w}}|}$, as a function of its minor radius $a$ (Panel a) and its axial length $L$ (Panel b), as listed in Table \ref{tab2}. The red lines display the fitting curves.}
\label{fig9}
\end{figure}

Figure \ref{fig8} displays the twist distribution from the MFR axis to its boundary in 4 cases with different $a$ and $R_{\rm C}$. It is seen that the twist of an MFR increases with its axial length but decreases with its minor radius. Interestingly, it is found that when $a$ is large or $R_{\rm C}$ is small, the twist becomes nearly uniformly distributed from the MFR axis to the boundary. Otherwise, the TDm flux ropes with the parabolic current density profile have a weakly twisted core and a highly twisted shell. Figure \ref{fig9} shows the relationship between the mean twist of an MFR and its minor radius in panel (a) and the axial length in panel (b). It is found in panel (a) that the mean twist of an MFR is inversely proportional to its minor radius with a fitting function of $|\overline{T_{\rm w}}| = 48.4(1/a) -0.095$ with a Pearson correlation coefficient of -0.920, where $a$ is in units of Mm. On the other hand, as shown in Figure \ref{fig9}b, the mean twist of an MFR increases linearly with its axial length, and the fitting function is $|\overline{T_{\rm w}}|=0.014L-0.48$ with a Pearson correlation coefficient of 0.996, where $L$ is in units of Mm. It is further found that the twist of an MFR is independent of the background potential field even though the magnetic flux $F$ of the MFR is different in those cases. The reason is explained as follows. For MFRs of the TDm model with the parabolic current profile, unlike the TD99 model \citep{td99}, the magnetic flux $F$ is a linear function of current $I$ governed by Equation (12) in \citet{Titov18}, so the twist of an MFR derived by the TDm model mainly depends on its geometry.

\begin{figure}[ht!]
\centering
\includegraphics[scale=0.8]{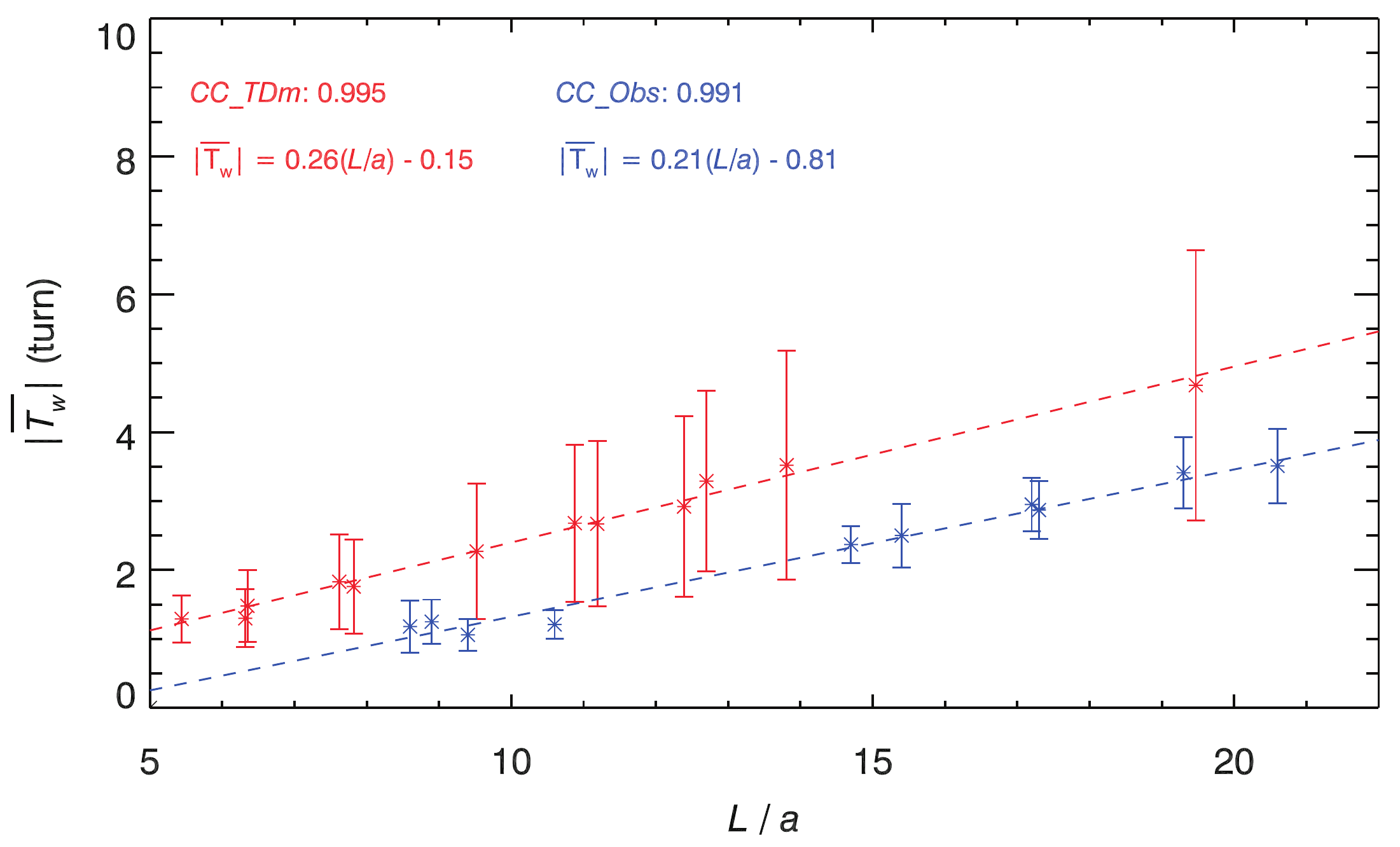}
\caption{Variation of the absolute value of the mean twist of the flux ropes versus their aspect ratio $L/a$. The red and blue asterisks represent the MFRs in the TDm theoretical models and observations, respectively, which are fit with the red and blue dashed lines, respectively.}
\label{fig10}
\end{figure}

Since the mean twist of an MFR is roughly proportional to both $1/a$ and $L$, it is straightforward to think that $|\overline{T_{\rm w}}|$ is proportional to the aspect ratio $L/a$ of the flux rope in the TDm model. We plot the variation of $|\overline{T_{\rm w}}|$ with $L/a$ as the red asterisks in Figure \ref{fig10}, where the data points are fit with the red dashed line. It is found that the variation can be fit with a linear function, i.e., $|\overline{T_{\rm w}}|=0.26(L/a)-0.15$, with a correlation coefficient of 0.995. In order to compare the $|\overline{T_{\rm w}}|$--$L/a$ relationships between observations and theoretical model, we overplot the observational results (i.e., Figure \ref{fig6}b) in Figure \ref{fig10}, as indicated by the blue asterisks along with the fitted line, i.e., the blue dashed line. It is found that, with the same aspect ratio $L/a$, the twist in observations is systematically smaller than that of TDm flux ropes by $\sim$1 turn, although their tendency is almost identical. This is explained as follows. The theoretical TDm flux ropes are closer to a force-free and divergence-free state, so generally only a few hundred iterations are needed in the magneto-frictional relaxation. However, to derive an MFR in a force-free state from observations, the magneto-frictional method is performed with much more iteration steps (according to our experience, about 60000 steps), during which the electric current density distribution inside the MFR keeps changing, reducing its twist. For example, the twist of an MFR decreases from 4.3 turns to 3.9 turns after relaxation in \citet{Guo19}. 

\section{Summary and Discussions}\label{sec:dis}

Twist is an important topological parameter of MFRs, and is closely related to its stability and dynamics. Studies showed that a flux rope is likely to become kink unstable when its twist exceeds a threshold. Additionally, assuming the magnetic helicity is conserved, some of the twist might be converted to writhe by an untwisting motion, which leads to the rotation of an MFR \citep{Green07}, and might significantly change their geoeffectiveness. Therefore, to estimate the twist of MFRs is an important topic in heliophysics. However, it is still difficult to obtain the twist directly so far, so we have to use some indirect methods. For example, the prominence material is a good tracer to the twisted magnetic structure. By analyzing the helical-threads and material motions in filaments, the twist of a flux rope can be estimated roughly \citep{Vrsnak91, Romano03}. In addition, if the feet of an MFR can be mapped, its twist can also be measured quantitatively according to the electric current distribution with the force-free assumption \citep[e.g.,][]{wangws19}. Another more widely used method is the NLFFF model. Then the twist can be calculated by the field lines geometry (formula (12) in \citet{Berger06}), or the parallel electric current (formula (16) in \citet{Berger06}) and these two metrics are close in the vicinity of the axis of the cylindrically symmetric magnetic configuration \citep{Liu16, Liu20}. However, these methods are either complex or limited by observations. Therefore, if we can understand how the twist of an MFR depends on its geometry and other routinely available magnetic properties (e.g., the aspect ratio and background magnetic field strength), we can estimate its twist in a more convenient way. In this paper, we used the RBSL technique to construct the force-free coronal magnetic field and to check how the magnetic twist changes with the input parameters.

We first constructed the magnetic structures for five filaments with the RBSL method and calculated their twists. Then we made a parameter survey for the TDm flux rope models with the parabolic electric current density distribution to explore how magnetic twist changes with the input parameters. According to our parameter survey, the twist of a force-free MFR derived by the RBSL method can be characterized by its aspect ratio, and is basically independent of the background magnetic field strength. In the cases of the observed filaments, the twist is related to the aspect ratio by $|\overline{T_{\rm w}}|=0.21(L/a)-0.81$; In contrast, in the cases of the TDm models, we have $|\overline{T_{\rm w}}|=0.26(L/a)-0.15$. On the one hand, such two quantitative relationships can help us estimate the mean twist of a force-free MFR. On the other hand, this result can deepen our understanding on the magnetic structures of solar filaments, coronal magnetic flux ropes, and interplanetary magnetic clouds.

\subsection{Difference of the magnetic configuration between two types of filaments}

Statistics shows that only 60\% of the active-region filaments are supported by flux ropes, whereas 96\% of the quiescent filaments are supported by flux ropes \citep{Ouyang17}. In this subsection, we try to provide an explanation for this observational result by investigating the factors affecting the twist of an MFR.

Our parameter survey indicates that the twist of an MFR is proportional to its axial length, inversely proportional to its minor radius, and basically independent of the background magnetic field strength. As mentioned before, there is much difference between active-region and quiescent filaments. For example, compared to active-region filaments, quiescent filaments are usually longer and lie along the weaker and dipole regions \citep{Tanberg-Hanssen1974, Tanberg1995, Filippov2000, Kuckein2009, Mackay2010, Engvold2015}, implying that quiescent filaments might have much longer magnetic field lines. Unfortunately, the minor radius of an MFR is difficult to be measured directly in observations. However, it is expected that active-region filaments with high magnetic twists and small minor radius are more prone to becoming kink unstable compared to quiescent filaments for two reasons. First, the resulting high magnetic pressure leads to a small plasma $\beta$ (the ratio of the gas pressure to magnetic pressure), hence a smaller threshold of kink instability according to \citet{Hood&Priest79}. Second, small minor radius means small volume of filament material. The resulting small weight of the filament cannot play the role to suppress kink instability as heavy filaments do \citep{Fan18}. Thus, we infer that the magnetic structures of quiescent filaments are likely to be more twisted than those of active-region filaments. This naturally accounts for why almost half of the active-region filaments are supported by sheared arcades, which are much less twisted compared to flux ropes.

\subsection{Existence of highly twisted flux ropes before eruption}

According to our results, the magnetic structures of longer filaments, e.g., quiescent filaments, would have large twist numbers, which are beyond the commonly mentioned threshold of kink instability. However, they survive our numerical simulations. Such a discrepancy raises a hotly debated question: Do highly twisted flux ropes exist in the solar corona?

With respect to the kink instability, the widely adopted critical value was proposed by \citet{Hood&Priest81}, where they considered a line-tying, force-free, uniform twisted flux rope and found that an MFR would become kink unstable when its twist exceeds 1.25 turns. However, both theories \citep{Hood&Priest79, Hood&Priest80, Bennett99, Baty01, Torok04, Torok05, Zaqarashvili10} and observations \citep{Wang16, liu19} demonstrated that the critical twist for the kink instability is not a constant. It depends on the external magnetic field \citep{Hood&Priest80, Torok05}, flux rope configuration \citep{Hood&Priest79,Baty01,Torok04}, plasma motions \citep{Zaqarashvili10}, plasma $\beta$ \citep{Hood&Priest79} and so on. Based on the previous works \citep{Dungey54, Hood&Priest79, Bennett99, Baty01}, the critical twist is proportional to the aspect ratio of an MFR, namely, $T_{\rm c}=\omega_{\rm c}L/(2\pi a)$, where $\omega_{\rm c}$ is a parameter related to the MFR structure. Thus, an MFR with a large aspect ratio might have a large critical twist for the kink instability. Besides, according to \citet{Hood&Priest79}, the critical twist increases with the plasma $\beta$. Compared to active-region filaments, quiescent filaments located in weaker magnetic regions have a larger plasma $\beta$, implying that the gas pressure contributes to make quiescent filaments more stable. Moreover, observations and numerical simulations also indicated that the gravity of a filament can suppress the onset of eruption, and the drainage can promote the eruption earlier \citep{Seaton11, Bi14, Jenkins18, Fan18, Fan20}. Therefore, the gravity of filaments should not be neglected when studying the stability of large quiescent filaments. In a word, with our results, we tend to believe that it is not impossible for some quiescent filaments to be stably supported by highly twisted MFRs with 3--5 turns even before eruption, as recently modeled by \citet{Mackay20}.

On the other hand, in many cases, a long quiescent filament is supported by several flux ropes or an arcade of weakly twisted (say, $\sim$1.5 turns) magnetic flux ropes as illustrated by the schematic Figure 6 in \citet{Chen2011}, rather than a single highly twisted flux rope. For example, \citet{Zheng17} reported an eruption event after the interaction between two filaments inside a filament channel. \citet{Luna17} reconstructed the coronal magnetic field for this event by using the flux-rope insertion method. It shows that the filament channel corresponds to two weakly twisted flux ropes or sheared arcades, as illustrated by Figure 13d in \citet{Luna17}.

Indeed, while some authors showed that large filaments can be supported by low-twist MFRs or sheared arcades \citep{Jibben16, Luna17}, others showed that large filaments can also be supported by a single highly-twisted MFR \citep{Su12, Su15, Mackay20}. Because of the intrinsic difficulty of coronal magnetic field extrapolation and the unreliable magnetic field measurement of the quiet region, the magnetic field and topological properties of quiescent filaments are still elusive. The specific magnetic structure supporting a quiescent filament might depend on the photospheric flows and their magnetic field evolution. We expect that future data-driven simulations can provide more hints.

\subsection{Origin of twists in interplanetary magnetic clouds}

Observations have revealed that large-scale interplanetary magnetic clouds (MCs) detected by insitu instruments usually have highly twisted magnetic structures. For example, \citet{Hu14} investigated 18 MCs with the Grad--Shafranov reconstruction method. They found that the mean twist of the MCs varies from 1.7 to 7.7 turns per AU, with an extreme case reaching 14.6 turns per AU. Besides, \citet{Farrugia99} studied a MC during 1995 October 24--25 with the uniform-twist GH model \citep{GH60} and found that the twist of this MC is about 8 turns per AU. Furthermore, \citet{Wang16} investigated 115 MCs measured at 1 AU with the velocity-modified-uniform-twist force-free flux rope model. It was found that most (80\%) of the MCs have a twist larger than 1.25 turns per AU, and some of the MCs have a twist larger than 10 turns per AU. Two factors are attributed to the high twist, one is the initial flux in the source region before eruption \citep{Xing20}, and the other is magnetic reconnection during eruption, which converts the ambient magnetic field to the MFR flux \citep{Longcope07, Qiu09, Aulanier12, Wangws17}. According to our results, quiescent filaments are generally more twisted than active-region filaments before eruption. It implies that for the highly twisted magnetic clouds originating from active regions, magnetic reconnection may contribute dominantly to the total twist, whereas for the magnetic clouds from quiet regions, most of the twist may come from the initial condition before eruption. Case studies could make this issue clearer. For example, for some active-region filament eruptions with major flares, events 4 and 6 in \citet{Hu14}, it is found that reconnection flux is even larger than the total flux of the magnetic cloud. However, for quiescent filament eruptions without major flares \citep[events 11 and 19 in][]{Hu14}, the reconnection flux is significantly smaller than the total flux of the magnetic cloud (note that some of the flux is lost due to erosion during propagation). We hope that statistical analyses with larger samples can give a sharper conclusion in the future.

\acknowledgments
The numerical calculations in this paper were performed in the cluster system of the High Performance Computing Center (HPCC) of Nanjing University. The AIA data are available courtesy of the SDO science team. The EUVI data are available courtesy of the STEREO/SECCHI consortium. This research was supported by NSFC (11961131002, 11533005, and 11773016), National Key Research and Development Program of China (2020YFC2201201), and the Science and Technology Development Fund of Macau, China (275/2017/A).

\bibliography{ms}{}
\bibliographystyle{aasjournal}

\newpage
\begin{deluxetable*}{lcccccccl}
\caption{A List of Five Filaments and Their Properties \label{tab1}}
\tablewidth{0pt}
\tablehead{
\colhead{Event time}& \colhead{Label} & \colhead{$L$$^{a}$} & \colhead{$a$$^{b}$} & \colhead{$L/a$} & \colhead{$|\overline{T_{\rm w}}|$$^{c}$} & \colhead{$\sigma_{J}$$^{d}$} &
\colhead{$\langle |f_{i}|\rangle$ $^{e}$} \\
\colhead{}& \colhead{}& \colhead{(Mm)} & \colhead{(Mm)} & \colhead{} & \colhead{(turn)}& \colhead{}  & \colhead{$(1.0\times 10^{-4})$}
}
\startdata
{2012-05-05-17:12:00 UT}& F1-1 & 145.0 & 16.2 & 8.9 & $1.25 \pm 0.32$ & 0.19 & 3.68& \\
{}& F1-2 & 145.0 & 17.0 & 8.6 & $1.18 \pm 0.38$ & 0.20 & 3.08& \\
\hline
{2010-08-07-16:00:00 UT}& F2-1 & 155.7 & 14.7 & 10.6 & $1.21 \pm 0.21$ & 0.31 & 3.60& \\
{}& F2-2 & 155.7 & 16.5 & 9.4 & $1.06 \pm 0.23$ & 0.29 & 2.97& \\
\hline
{2011-06-21-01:12:26 UT}& F3-1 & 318.2 & 15.5 & 20.6 & $3.51 \pm 0.54$ & 0.36 & 1.67& \\
{}& F3-2 & 318.2 & 20.7 & 15.4 & $2.50 \pm 0.46$ & 0.34 & 1.31& \\
\hline
{2011-07-08-21:36:00 UT}& F4-1 & 330.5 & 17.1 & 19.3 & $3.41 \pm 0.52$ & 0.20 & 2.38& \\
{}& F4-2 & 330.5 & 19.2 & 17.2 & $2.95 \pm 0.39$ & 0.23 & 2.54& \\
\hline
{2012-05-10-00:00:00 UT}& F5-1 & 260.2 & 15.0 & 17.3 & $2.87 \pm 0.42$ & 0.27 & 1.67& \\
{}& F5-2 & 260.2 & 17.7 & 14.7 & $2.37 \pm 0.27$ & 0.28 & 1.64& \\
\hline
\enddata
\tablecomments{$^{\rm a}$ The axis length of an MFR.\\$^{\rm b}$The minor radius of an MFR defined by QSLs.\\$^{\rm c}$$|\overline{T_{\rm w}}|$ denotes the absolute value of the mean twist of a flux rope. The error bar is the standard deviation of all selected field lines.\\$^{\rm d}$force-free metric.\\$^{\rm e}$divergence-free metric.}
\end{deluxetable*}

\begin{deluxetable*}{lcccccccccccl}
\caption{TDm models and Their Properties\label{tab2}}
\tablehead{
\colhead{}& \colhead{Case} & \colhead{$R_{\rm c}$$^{a}$}& \colhead{$L$} & \colhead{$a$}& \colhead{$L/a$} & \colhead{$d_{q}$$^{b}$} & \colhead{$L_{q}$$^{c}$}& \colhead{$q$$^{d}$} & \colhead{$|\overline{T_{\rm w}}|$} & \colhead{$\sigma_{J}$} &
\colhead{$\langle |f_i|\rangle$} \\
\colhead{}& \colhead{}& \colhead{(Mm)} & \colhead{(Mm)}& \colhead{(Mm)}& \colhead{} & \colhead{(Mm)} & \colhead{(Mm)} & \colhead{$\rm (T \cdot Mm^{2})$} & \colhead{(turn)}& \colhead{}  & \colhead{$(1.0\times 10^{-4})$}
}
\startdata
{}& A1 & 94.2 & 190.5 & 10 & 19.5 & 50 & 50 & 100 & $4.68 \pm 1.96$ & 0.086 & 0.072 &\\
{}& A2 & 94.2 & 190.5 & 15 & 12.7 & 50 & 50 & 100 & $3.29 \pm 1.31$ & 0.075 & 0.014 & \\
{}& A3 & 94.2 & 190.5  & 17.5 & 10.9 & 50 & 50 & 100 & $2.68 \pm 1.14$ & 0.089 & 0.078 & \\
{$a$}& RF & 94.2 & 190.5  & 20 & 9.5 & 50 & 50 & 100 & $2.27 \pm 0.98$ & 0.086 & 0.147 & \\
{}& A4 & 94.2 & 190.5  & 25 & 7.6 & 50 & 50 & 100 & $1.83 \pm 0.69$ & 0.089 & 0.134 & \\
{}& A5 & 94.2 & 190.5  & 30 & 6.4 & 50 & 50 & 100 & $1.48 \pm 0.52$ & 0.096 & 0.157 & \\
{}& A6 & 94.2 & 190.5  & 35 & 5.4 & 50 & 50 & 100 & $1.29 \pm 0.34$ & 0.100 & 0.186 & \\
{}& A7 & 94.2 & 190.5  & 40 & 4.8 & 50 & 50 & 100 & $1.12 \pm 0.28$ & 0.104 & 0.245 & \\
\hline
{}& R1 & 75.1 & 126.4 & 20 & 6.3 & 50 & 50 & 100 & $1.30 \pm 0.42$ & 0.138 & 0.094 & \\
{}& R2 & 83.9 & 156.4 & 20 & 7.8 & 50 & 50 & 100 & $1.76 \pm 0.68$ & 0.139 & 0.094 & \\
{$R_{c}$}& RF & 94.2 & 190.5 & 20 & 9.5 & 50 & 50 & 100 & $2.27 \pm 0.98$ & 0.086 & 0.074 & \\
{}& R4 & 104.5 & 223.7 & 20 & 11.2 & 50 & 50 & 100 & $2.67 \pm 1.20$ & 0.078 & 0.147 & \\
{}& R5 & 111.9 & 247.8 & 20 & 12.4 & 50 & 50 & 100 & $2.92 \pm 1.31$ & 0.069 & 0.134 & \\
{}& R6 & 120.8 & 276.1 & 20 & 13.8 & 50 & 50 & 100 & $3.52 \pm 1.66$ & 0.062 & 0.104 & \\
\hline
{}& B1 & 94.2 & 190.5  & 20 & 9.5 & 50 & 50 & 50 & $2.28 \pm 0.97$ & 0.085 & 0.148 & \\
{}& RF & 94.2 & 190.5  & 20 & 9.5 & 50 & 50 & 100 & $2.29 \pm 0.96$ & 0.086 & 0.147 & \\
{$q$}& B2 & 94.2 & 190.5  & 20 & 9.5 & 50 & 50 & 150 & $2.29 \pm 0.96$ & 0.087 & 0.146 & \\
{}& B3 & 94.2 & 190.5  & 20 & 9.5 & 50 & 50 & 200 & $2.28 \pm 0.98$ & 0.085 & 0.148 & \\
{}& B4 & 94.2 & 190.5  & 20 & 9.5 & 50 & 50 & 250 & $2.29 \pm 0.96$ & 0.087 & 0.147 & \\
{}& B5 & 94.2 & 190.5  & 20 & 9.5 & 50 & 50 & 300 & $2.30 \pm 0.96$ & 0.088 & 0.147 & \\
\hline
\enddata
\tablecomments{$^{\rm a}$ The major radius of the  MFR.\\$^{\rm b}$The depth of two magnetic charges.\\$^{\rm c}$The distance from the magnetic charge to the plane $y=0$.\\$^{\rm d}$ Magnetic charge strength.\\ The other notations are the same as in Table \ref{tab1}.}
\label{tab2}
\end{deluxetable*}

\end{document}